\documentclass[preprint,12pt,nofootinbib,groupedaddress,superscriptaddress]{revtex4-1}

\usepackage{graphicx}
\usepackage{dcolumn} 
\usepackage{bm}
\usepackage{amssymb}
\usepackage{amsmath}
\usepackage{epsfig}    
\usepackage{color}
\usepackage{slashed}
\usepackage{hyperref} 
\usepackage{comment}

\begin{document}

\title{Multi-Lepton Jets from Quadruple $Z'$ via the Higgs Decay at LHC}
\preprint{OU-HET-1254}

\author{Jinmian Li}
\email{jmli@scu.edu.cn}
\affiliation{College of Physics, Sichuan University, Chengdu 610065, China}

\author{Takaaki Nomura}
\email{nomura@scu.edu.cn}
\affiliation{College of Physics, Sichuan University, Chengdu 610065, China}

\author{Kei Yagyu}
\email{yagyu@het.phys.sci.osaka-u.ac.jp}
\affiliation{Department of Physics, Osaka University, Toyonaka, Osaka 560-0043, Japan}

\begin{abstract}

\noindent
We investigate multi-lepton jet events from the decay of the 125 GeV Higgs boson ($h$) into quadruple new gauge bosons $(Z')$ at the LHC. 
Such an exotic decay is realized via the process of $h \to \phi \phi \to Z'Z'Z'Z'$ with new scalar boson $\phi$ in models with an additional $U(1)$ gauge symmetry. Charged leptons coming from the $Z'$ decay tend to be observed as lepton-jets rather than isolated leptons when the masses of $Z'$ and $\phi$ are smaller than ${\cal O}$(10) GeV, because of the highly-boosted effects. 
Performing the signal and background analyses, we find that the branching ratio of $h \to 4Z'$ is maximally constrained to be smaller than of order $10^{-6}$ ($10^{-7}$) by using the muonic-lepton jets assuming the integrated luminosity of 140 fb$^{-1}$ (3000 fb$^{-1}$) at LHC. 
For lighter $Z'$ ($< 2m_\mu$), we can use the electronic-lepton jets instead of the muon-jets, by which the upper limit on the branching ratio is obtained to be of order $10^{-6}$-$10^{-5}$.
These bounds can be converted into the constraint on model parameters such as a mixing angle between $h$ and $\phi$. It is shown that stronger bounds on the mixing angle are obtained in the dark photon case as compared with the previous constraints given by flavor experiments and the Higgs decay $h \to Z'Z'$ in the mass range of $m_{Z'}\lesssim 10$ GeV.

\end{abstract}

\maketitle
\newpage


\section{Introduction}

The discovery of the new neutral gauge boson $Z'$ is direct evidence of the existence of a fifth fundamental force in nature.
Such a new force appears in various new physics beyond the Standard Model (SM) including an extra $U(1)$ gauge symmetry, e.g., grand unified theories, models explaining neutrino masses, dark matter and baryon asymmetry of the Universe. 
Searches for $Z'$ have been done by various experiments, e.g., collider experiments, flavor experiments, and cosmological observations. 
For instance, searches for a new dilepton resonance have been performed at the LHC, where a high-mass range of $Z'$, 250 GeV $<m_{Z'} <$ 6 TeV, has been surveyed, and provided the constraint $m_{Z'}\gtrsim 5$ TeV for $Z'$ having couplings to SM fermions with a similar size as those of the SM $Z$ boson~\cite{ATLAS:2019erb}.  
A similar search has been performed for the mass range 11.5 GeV $< m_{Z'} <$ 200 GeV, which has given the upper limit on the kinetic mixing parameter $\epsilon$ to be around $10^{-2}$-$10^{-3}$ in the dark photon case~\cite{CMS:2019buh}.  
For a lighter $Z'$ ($200~\text{MeV}\lesssim m_{Z'}\lesssim 10$ GeV), flavor experiments have played an important role in putting an upper limit on $\epsilon$, typically $\epsilon < {\cal O}(10^{-3}\text{-}10^{-4})$ at BaBar~\cite{BaBar:2014zli,BaBar:2016sci} and LHCb~\cite{LHCb:2017trq,LHCb:2019vmc}.   
An even lighter $Z'$ tends to be long-lived, so that beam dump experiments are useful to search for such $Z'$ with $\sim 100$ MeV~\cite{Riordan:1987aw,Bjorken:1988as,Batell:2014mga,Marsicano:2018krp,Blumlein:2011mv,Blumlein:2013cua,Gninenko:2012eq}.   

In addition to the above searches, it is quite useful to consider the decays of the discovered Higgs boson ($h$) and $Z$ boson into the final states including $Z'$ as long as they are kinematically allowed. 
These searches can be performed at Higgs factories, e.g., the High-Luminosity LHC (HL-LHC)~\cite{ATLAS:2013hta,CMS:2013xfa}, the International Linear Collider (ILC)~\cite{ILC:2013jhg,ILCInternationalDevelopmentTeam:2022izu}, the Circular Electron-Positron Collider (CEPC)~\cite{CEPCStudyGroup:2018ghi}, the Future Circular Collider (FCC-ee)~\cite{FCC:2018byv} and the Compact LInear Collider (CLIC)~\cite{CLIC:2016zwp}, as well as $Z$ factories, e.g., the CEPC~\cite{CEPC-SPPCStudyGroup:2015csa} and the FCC-ee~\cite{FCC:2018byv}.  
Recently in Ref.~\cite{Nomura:2024bsz}, triple $Z'$ signatures 
have been proposed via the decay of $Z \to Z'\phi \to Z'Z'Z'$ with $\phi$ being a new scalar boson whose Vacuum Expectation Value (VEV) gives the mass of $Z'$, by which the $\epsilon$ parameter can significantly be constrained to be smaller than $10^{-6}$ in the dark photon case.  
The Higgs decays into multi-$Z'$ bosons have also been discussed in Refs.~\cite{Gopalakrishna:2008dv,Davoudiasl:2013aya,Chang:2013lfa,Curtin:2013fra,Falkowski:2014ffa,Curtin:2014cca,Nomura:2024pwr} for $h \to Z'Z'$
and in Refs.~\cite{Chang:2013lfa,Izaguirre:2018atq,Nomura:2024pwr} for $h \to 4Z'$. The former process has already been surveyed at LHC~\cite{CMS:2021pcy,ATLAS:2021ldb}, and the upper limit on the branching ratio of the Higgs boson has been taken, which can be converted into the constraint on model parameters such as the mixing angle between $h$ and $\phi$ and the new $U(1)$ gauge coupling~\cite{Nomura:2024pwr}. 

In this paper, we investigate the Higgs decays into four $Z'$, i.e., $h \to \phi\phi \to Z'Z'Z'Z'$, where the masses of $Z'$ and $\phi$ are much smaller than that of $h$. 
In this case, produced $Z'$ and $\phi$ can highly be boosted, and thus some of the charged leptons generated via the decay of $Z'$ can highly be collimated. This indicates that charged leptons in the final states can be observed as ``Lepton-Jets (LJs)" rather than isolated leptons. 
In fact, the searches for LJs have already been performed at LHC, in which the target signal is supposed to be $h \to Z'Z'$ with a collimated pair of muons or electrons from the $Z'$ decay~\cite{ATLAS:2024zxk,CMS:2024jyb}. 
The upper limit on the branching ratio of $h \to Z'Z'$ have been taken to be of order $10^{-3}$-$10^{-2}$ ($10^{-5}$-$10^{-4}$) using the signal with an electron (muon) pair. 

We investigate the multi-LJs signatures coming from the Higgs decay $h \to \phi\phi \to Z'Z'Z'Z'$ with the decay of $Z' \to \ell^+\ell^-$ ($\ell=e,\mu$). 
We perform the simulation study at the LHC including detector-level effects, 
and show the distribution of some observables such as the number of isolated leptons, the invariant mass of each LJ, and the invariant mass of the LJ pair. 
We find that most of the background events can be eliminated by requiring that the final states contain at least two LJs. Furthermore, in most cases, the signal significance is maximized by requiring that each LJ contains at least three charged leptons.  
We then demonstrate the upper limit on the branching ratio of the Higgs boson by imposing the optimized kinematical cuts, which is converted into the constraint on the upper limit on a mixing parameter between $h$ and $\phi$.  

This paper is organized as follows. 
In Sec.~\ref{sec:2}, we give a brief review on models with an additional $U(1)$ gauge symmetry which induces $Z'$ in a simple framework. 
In Sec.~\ref{sec:3}, we discuss the reconstruction of the LJs, and define our kinematical cuts in order to extract the signal events from the backgrounds. We then give an upper limit on the branching ratio of the $h \to 4Z'$ mode. 
Sec.~\ref{sec:4} is devoted to demonstrating the constraint on the model parameters by using the upper limits on the branching ratio of the Higgs boson given by the 4$Z'$ events. 
Conclusions are given in Sec.~\ref{sec:5}.

\section{Model \label{sec:2}}

We briefly review models with an extra $U(1)_X$ gauge symmetry which is spontaneously broken by a complex singlet scalar field $\Phi$ with a non-zero $U(1)_X$ charge. 

The Lagrangian for the purely bosonic sector is given as follows: 
\begin{align}
{\cal L}  & \supset  
|D_\mu H|^2 +  |D_\mu \Phi|^2 
-\frac{1}{4} B_{\mu \nu} B^{\mu \nu} -\frac{1}{4} X_{\mu \nu} X^{\mu \nu}  - V(H,\Phi), \label{eq:lagrangian}
\end{align}
where $H$ and $B_{\mu\nu}$ [$X_{\mu\nu}$] denote the Higgs doublet field and the field strength tensor for $U(1)_Y$  $[U(1)_X]$ gauge bosons $B_\mu$ [$X_\mu$], respectively. 
These gauge bosons are defined in the basis where the kinetic mixing $\epsilon$ between $B_\mu$ and $X_\mu$ is eliminated by the non-unitary transformation. 
As we will see soon below, the effect of $\epsilon$ appears in the covariant derivative with respect to the $X_\mu$ field, and it leads to the $Z$ and $Z'$ boson mixing. 

In Eq.~(\ref{eq:lagrangian}), $V(H,\Phi)$ represents the Higgs potential, and its most general form is given by 
\begin{align}
V = & -\mu_H^2 |H|^2 -\mu_\Phi^2|\Phi|^2 + \frac{\lambda_H}{2}|H|^4 
+ \frac{\lambda_\Phi}{2}|\Phi|^4 + \lambda_{H\Phi}|H|^2|\Phi|^2, 
\end{align}
where the scalar fields are parameterized as 
\begin{align}
H = \begin{pmatrix}
G^+ \\
\frac{1}{\sqrt{2}}(\tilde{h} + v + iG)
\end{pmatrix},~~
\Phi = \frac{1}{\sqrt{2}} (\tilde{\phi} + v_\Phi^{} + iG'), 
\end{align}
with $v$ and $v_\Phi^{}$ being the VEVs of $H$ and $\Phi$, respectively, and $G^\pm$, $G$ and $G'$ being the Nambu-Goldstone bosons which are absorbed into $W^\pm$, $Z$ and $Z'$ bosons, respectively.  
The component fields $\tilde{h}$ and $\tilde{\phi}$ are mixed via the $\lambda_{H\Phi}$ term as follows: 
\begin{align}
\begin{pmatrix}
\tilde{h} \\
\tilde{\phi}
\end{pmatrix}=
R(\alpha)
\begin{pmatrix}
h \\
\phi
\end{pmatrix},~~
R(\theta) \equiv 
\begin{pmatrix}
\cos\theta & -\sin\theta \\
\sin\theta & \cos\theta
\end{pmatrix},
\end{align}
with the mixing angle $\alpha$ given by
\begin{align}
\tan2\alpha = \frac{2\lambda_{H\Phi}vv_\Phi}{\lambda_H v^2- \lambda_\Phi v_\Phi^2}. 
\end{align}

The covariant derivative $D_\mu$ for a field $\Psi$ can be expressed as 
\begin{align}
D_\mu\Psi  \supset & \left[\partial_\mu  -ieQ_\Psi A_\mu
-ig_Z^{}(T_\Psi^3 - \sin^2\theta_W Q_\Psi)\tilde{Z}_\mu -ig_X^{}X_\Psi X_\mu \right]\Psi, \label{eq:cov}
\end{align}
where $Q_\Psi$, $T_\Psi^3$ and $X_\Psi$ respectively represent 
the electric charge, the third component of the isospin and the {\it effective} $U(1)_X$ charge for $\Psi$, while $\theta_W$ is the Weinberg angle, $g_X$ is the gauge coupling of $U(1)_X$ and $g_Z =g/\cos\theta_W^{}$ with $g$ being the $SU(2)_L$ gauge coupling. 
The last one is given by the original $U(1)_X$ charge $\tilde{X}_\Psi$ and the contribution from the kinetic mixing $\epsilon$ as follows:
\begin{align}
X_\Psi &= \tilde{X}_\Psi - Y_\Psi\frac{g}{g_X^{}}\tan\theta_W\tan\epsilon, \label{eq:x-charge}
\end{align}
where $Y_\Psi$ is the hypercharge. 
In Eq.~(\ref{eq:cov}), the hypercharge field $B_\mu$ is converted into $B_\mu = \cos\theta_W A_\mu - \sin\theta_W \tilde Z_\mu$ with $A_\mu$ being the massless photon field. 
The neutral gauge bosons 
$\tilde{Z}_\mu$ and $X_\mu$
can be mixed as 
\begin{align}
\begin{pmatrix}
\tilde{Z}_\mu\\
X_\mu
\end{pmatrix}=
R(\zeta)
\begin{pmatrix}
Z_\mu\\
Z'_\mu
\end{pmatrix}, ~~
\end{align}
where the mixing angle $\zeta$ is given by 
\begin{align}
\sin 2 \zeta = \frac{g_Z^{}g_X^{}  X_H v^2}{m^2_{Z'} - m^2_Z}. 
\end{align}
We identify $Z_\mu$ and $Z_\mu'$ with the SM $Z$ boson and the additional $Z'$ boson, respectively. 

The Higgs boson $h$ can decay into a pair of $Z'$ and $\phi$ through the mixing with $\phi$. These decay rates are given by 
\begin{align}
\Gamma(h \to Z' Z') &= \frac{ m_h^3\sin^2 \alpha}{32 \pi v_\Phi^2 } ( 1 - 4x_{Z'} + 12x_{Z'}^2 )\beta(x_{Z'}), \label{eq:hzpzp}\\
\Gamma(h \to \phi \phi) &= \frac{m_h^3\sin^2\alpha}{32 \pi v_\Phi^2} \cos^2\alpha(1 + 2x_\phi)\left(\cos\alpha + \sin\alpha\frac{v_\Phi}{v}\right)^2
\beta(x_\phi), \label{decay:hphiphi} 
\end{align} 
where $x_i = m_i^2/m_h^2$ and $\beta(x) = \sqrt{1 - 4x}$. 
In Eq.~(\ref{eq:hzpzp}), 
we drop the contribution from the term proportional to $\zeta^2(m_{Z'}/m_Z)^2\sin\alpha$ which is induced by the interference between the $Z$-$Z'$ mixing and the Higgs mixing. 
This contribution is negligibly smaller than the dominant contribution simply coming from the Higgs mixing $\sin^2\alpha$ in our scenario. 
We also note that two decay rates of $h\to Z'Z'$ and $h \to \phi\phi$ become comparable when $x_{Z'},~x_{\phi},~|\sin\alpha| \ll 1$~\cite{Nomura:2024pwr}. 

The new Higgs boson $\phi$ can decay into a pair of SM particles and a pair of $Z'$. 
The decay rate of the former is simply given by multiplying $\sin^2\alpha$ to the corresponding expression in the SM with the replacement of $m_h \to m_\phi$, 
while the latter is given by 
\begin{align}
\Gamma(\phi \to Z'Z') &= \frac{ m_\phi^3\cos^2 \alpha}{32 \pi v_\Phi^2 } ( 1 - 4y_{Z'} + 12y_{Z'}^2 )\beta(y_{Z'}), \label{eq:phi-decay-zpzp}
\end{align}
with $y_{Z'} = m_{Z'}^2/m_\phi^2$. 
Again, we neglect the contribution from $\zeta^2(m_{Z'}/m_Z)^2\sin\alpha$. 

\section{Simulation studies for multi-LJs\label{sec:3}}

\begin{figure}
    \centering
    \includegraphics[width=0.5\linewidth]{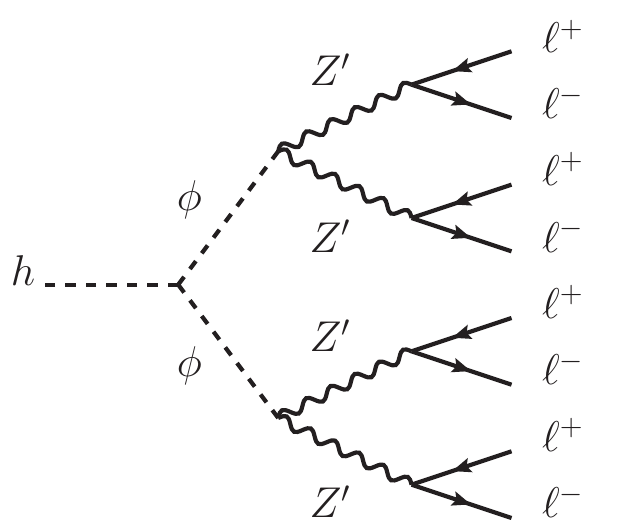}
    \caption{Exotic decay of the 125 GeV Higgs boson with eight charged leptons in the final state ($\ell = e$ or $\mu$). Some of these charged leptons can form LJs. }
    \label{fig:diagram}
\end{figure}

\subsection{LJs from the signal events}

We consider the exotic decay of the SM-like Higgs boson $h \to \phi \phi \to (Z^\prime Z^\prime) (Z^\prime Z^\prime)$ at the 13 TeV LHC with $Z^\prime$ subsequently decaying into either an electron or a muon pair, see Fig.~\ref{fig:diagram}. 
As we mentioned in the Introduction, some of the charged leptons are not observed as isolated leptons but LJs if the masses of $Z'$ and $\phi$ are sufficiently smaller than the Higgs boson mass.
LJs are typically defined by requiring two or more charged leptons found in a region with $\Delta R <0.4$~\cite{ATLAS:2024zxk}, where the angular distance $\Delta R$ is defined by $\Delta R \equiv \sqrt{\Delta \eta^2 +\Delta \phi^2}$ with 
$\Delta \eta$ and $\Delta \phi$ being a difference of the pseudo-rapidity and azimuthal angle, respectively, between two particles. 
The $\Delta R$ value measures the cone size of the LJ formed by the decay of a boosted $\phi$ into four lepton final states. 

\begin{figure}[htbp]
	\begin{center}
		\includegraphics[width=0.6\textwidth]{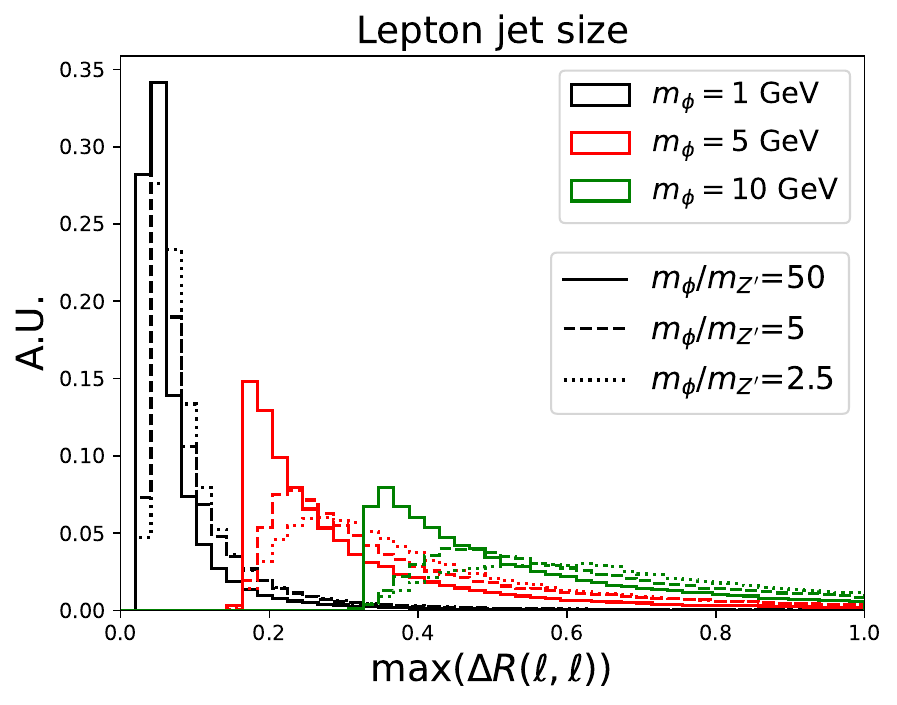}
	\end{center}
\caption{The distributions of the $\max(\Delta R(\ell,\ell))$ in a $\phi$ decay, for the decay channel $h \to \phi \phi \to (Z^\prime Z^\prime) (Z^\prime Z^\prime) \to (4\ell) (4\ell)$. The masses of $\phi$ and $Z^\prime$ are indicated in the legends. } 
\label{fig:drs4}
\end{figure}

In order to demonstrate how smaller $\Delta R$ is obtained from the decay of the Higgs boson, 
we plot the distributions of $\max[\Delta R(\ell,\ell)]$ at parton level in Fig.~\ref{fig:drs4}, where we take into account all the possible six combinations of two charged leptons in a $\phi$ decay. 
We can observe that the lower cutoff on the LJ size is given by $R_{\text{cut}}=(2 \times m_{\phi}) / (m_{h}/2)$~\cite{Butterworth:2008iy}, where $m_{h}/2$ in the denominator refers to the typical energy of the $\phi$ in the production process. 
The shape of $\max[\Delta R(\ell,\ell)]$ is mainly determined by the mass ratio $m_{\phi}/m_{Z^\prime}$, e.g., a higher mass ratio renders a lepton pair from $Z^\prime$ more collimated. As a result, the cone size of the LJ is more concentrated at around the $R_{\text{cut}}$. 
We see that most of the lepton pair takes $\Delta R$ to be smaller than 0.4 for $m_{Z'},~m_{\phi}$ to be smaller than about 10 GeV. 

\subsection{Signal and background simulations}

The signal events are simulated within the \texttt{MadGraph5\_aMC@NLO}~\cite{Alwall:2014hca} framework which uses the UFO model files for the model in Sec.~\ref{sec:2} generated by \texttt{FeynRules}~\cite{Alloul:2013bka}. 
The effects of initial state radiation and final state radiation, as well as the hadronization and decay of the SM hadrons, are simulated by the \texttt{Pythia8}~\cite{Sjostrand:2007gs}. 
A fast simulation of detector responses is performed by  \texttt{Delphes}~\cite{deFavereau:2013fsa} with the default ATLAS configuration card. 

We will follow the procedures adopted in the ATLAS analysis~\cite{ATLAS:2024zxk} to reconstruct the LJs. 
We first impose the following basic kinematical cuts for each electron or muon:
\begin{align}
&p_T^e \geq 4.5~\text{GeV},~~|\eta^e| \leq 2.47,\\
&p_T^\mu \geq 3~\text{GeV},~~|\eta^\mu| \leq 2.5,
\end{align}
with $p_T^\ell$ and $\eta^\ell$ being the transverse momentum and pseudo rapidity for $\ell$, respectively. 

Second, in the final state of simulated events, we identify isolated electrons and muons following the calorimeter-based loose criterion defined in Ref.~\cite{ATLAS:2019qmc} and the PFlowLoose isolation requirement defined in Ref.~\cite{ATLAS:2020auj}, respectively. 
Then we apply the Cambridge–Aachen clustering~\cite{Dokshitzer:1997in} with a radius parameter $R=0.4$ and minimal transverse momentum of jet $p_T^{\min}=6$ GeV to reconstruct the (lepton-)jets. 
To avoid double-counting of lepton and (lepton-)jets, the reconstructed jets are discarded if they are found within a $\Delta R = 0.2$ cone around an isolated lepton. Moreover, the remaining jets are retained against isolated leptons if they satisfy $\Delta R (\ell ,j) < \min(0.4,0.04+10~\text{GeV}/p_T^{\ell}) $. 

Finally, we identify the LJs by requiring that a jet should contain at least two lepton constituents.
However, in the case of an electronic LJ, electromagnetic showers from two close-by electrons will be merged into a single cluster, which is reconstructed as a single electron with two associated tracks.
Thus, an inner-detector (ID) track (corresponding to a charged hadron), which is in a $\Delta R = 0.15$ cone around an electron, will be also considered as a misidentified electron in LJ recognition. 
Jets with less than 2 (misidentified-)leptons are recognized as hadronic jets.  

As discussed in Ref.~\cite{ATLAS:2019qmc}, the main background for the electronic LJs comes from the random overlap of electrons, produced in $Z$ boson decays, with additional ID tracks. While the background for the muonic LJs mainly from the overlap of muons in $Z$ boson and hadron decays.
We simulate the $Z_\ell$ + jets background with up to two additional hadronic jets at parton level, where $Z_\ell$ denotes that the $Z$ boson decaying into either an electron or a muon pair. The MLM matching procedure~\cite{Alwall:2007fs} is adopted to avoid double counting between the matrix element calculation and the parton showering. 
The other dominant background comes from the $Z_\ell Z_\ell$ process.
The subdominant backgrounds are attributed to QCD processes with muons from bottom quark decay and vector meson resonances~\cite{CMS:2024jyb}, which we will not consider in this work. 

Candidate signal events are required to contain at least two LJs. For events with more than two LJ candidates, the analysis selects the leading LJ in $p_T$ and the LJ with the largest azimuthal angle ($\Delta \phi$) separation from the leading LJ.
To investigate the impact of varying the number of lepton constituents requirements within LJs, three signal regions (SRs) are defined:
\begin{itemize}
\item[\textbf{SR1}:] Both selected LJs must contain at least three charged leptons.
\item[\textbf{SR2}:] One selected LJ must contain at least three charged leptons, while the other must contain at least two charged leptons.
\item[\textbf{SR3}:] Only basic LJ requirement is applied, that is both selected LJs must contain at least two charged leptons.
\end{itemize}

In each SR, we require 
$m_{\text{LJ1,LJ2}}^{} < 130$ GeV, where $m_{\text{LJ1,LJ2}}^{}$
denotes the invariant mass of the LJ pair, i.e., the reconstructed Higgs mass, 
which is defined as 
\begin{align}
m_{\text{LJ1,LJ2}}^{} =  \sqrt{(p_{\rm LJ1}^\mu + p_{\rm LJ2}^\mu)^2}, 
\end{align}
where $p^\mu_{\rm LJ}=\sum_i p^{\mu}_i$ with $i$ running over all leptonic constituents.
%
Moreover, when $m_{Z^\prime} \gtrsim 2 m_\mu$, only the LJs that contain at least two muons are considered, as the electronic LJs suffer from the serious background where the ID tracks mimic the electrons. 
In the parameter space $m_{Z^\prime} < 2 m_\mu$, only the electronic LJs exist in the signal process. We find that applying the invariant mass cut on the LJ is quite efficient in suppressing the backgrounds.
Considering the detector resolution, we require the invariant mass of the LJs
($m_{\text{LJ}} = \sqrt{(p_{\rm LJ}^\mu)^2}$)
to be either smaller than 1 GeV or 2 GeV. The one that provides higher signal significance will be used to calculate the bound at given $m_\phi$ and $m_{Z^\prime}$.

\begin{table}[tbp]
	\centering 
	\begin{tabular}{|c|c|c|} \hline
   & $Z_\ell Z_\ell$  & $Z_\ell$ + jets \\ \hline
  SR1-$e$  & $6.8\times 10^{-6}$ ($1.2\times 10^{-6}$) [$\lesssim 4\times 10^{-7}$] & $2.31\times 10^{-6}$ ($5.53 \times 10^{-7}$) [$5.03 \times 10^{-8}$]  \\ \hline
  SR2-$e$ & $1.44 \times 10^{-5}$ ($2.4 \times 10^{-6}$) [$4 \times 10^{-7}$] & $6.39 \times 10^{-6}$ ($2.26 \times 10^{-6}$) [$1.51 \times 10^{-7}$] \\ \hline
  SR3-$e$ & $2.88 \times 10^{-5}$ ($6.4 \times 10^{-6}$) [$8 \times 10^{-7}$] &  $9.91 \times 10^{-6}$ ($4.93 \times 10^{-6}$) [$1.11 \times 10^{-6}$] \\ \hline \hline
  SR1-$\mu$ &  $\lesssim 4 \times 10^{-7}$ &  $\lesssim 5.03 \times 10^{-8}$ \\ \hline
  SR2-$\mu$ & $\lesssim 4 \times 10^{-7}$ &  $1.01 \times 10^{-7}$  \\\hline
  SR3-$\mu$ &  $2.28 \times 10^{-5}$ & $4.52 \times 10^{-7}$  \\\hline
	\end{tabular}
	\caption{\label{tab:efficiency}
    Efficiencies for different SRs for the $Z_\ell Z_\ell$ and $Z_\ell$+jets backgrounds. Values in the parenthesis and square brackets correspond to the efficiencies with invariant mass cuts on electronic LJs $m_{\text{LJ}}<2$ GeV and  $m_{\text{LJ}}<1$ GeV, respectively. }
\end{table}

In Table~\ref{tab:efficiency}, we present the selection efficiencies of three SRs for $Z_\ell Z_\ell$ and $Z_\ell$ + jets backgrounds. We can find that for the muonic LJ case, demanding both LJs to have at least three lepton constituents (at least two of them should be muon) can help suppress both backgrounds to a negligible level. 
In contrast, the electronic LJ case exhibits significantly higher background levels due to the challenges of distinguishing electronic LJs from the overlapping of an electron and ID tracks.
The invariant mass cut on the LJ is essential in suppressing both the $Z_\ell Z_\ell$ and the $Z_\ell$+jets backgrounds. In particular, requiring $m_{\text{LJ}}<1$ GeV can reduce the background to a similar amount as that in the muonic channel for the SR1.

\begin{figure}[t]
	\begin{center}
		\includegraphics[width=0.32\textwidth]{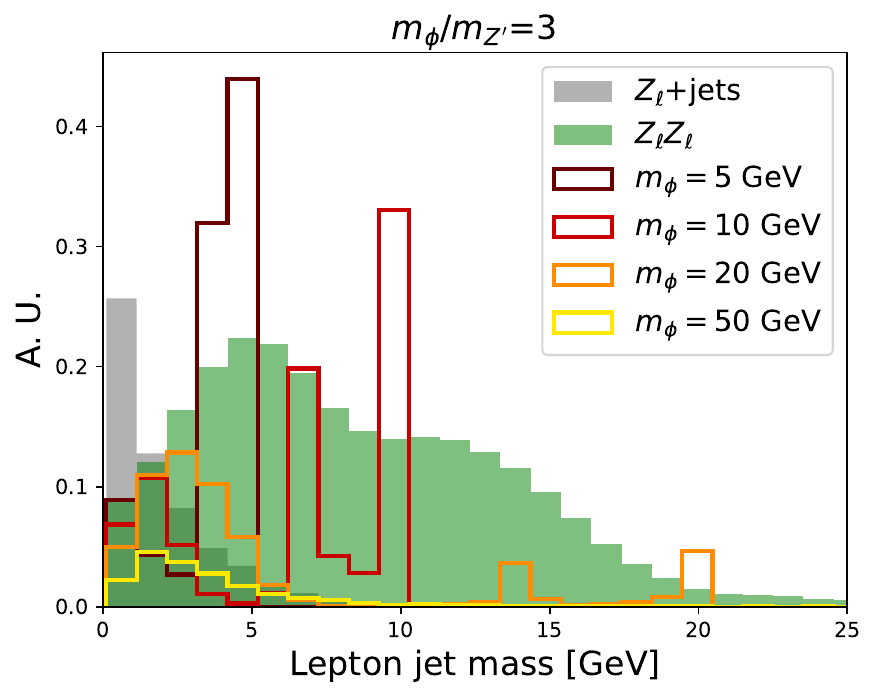}
		\includegraphics[width=0.32\textwidth]{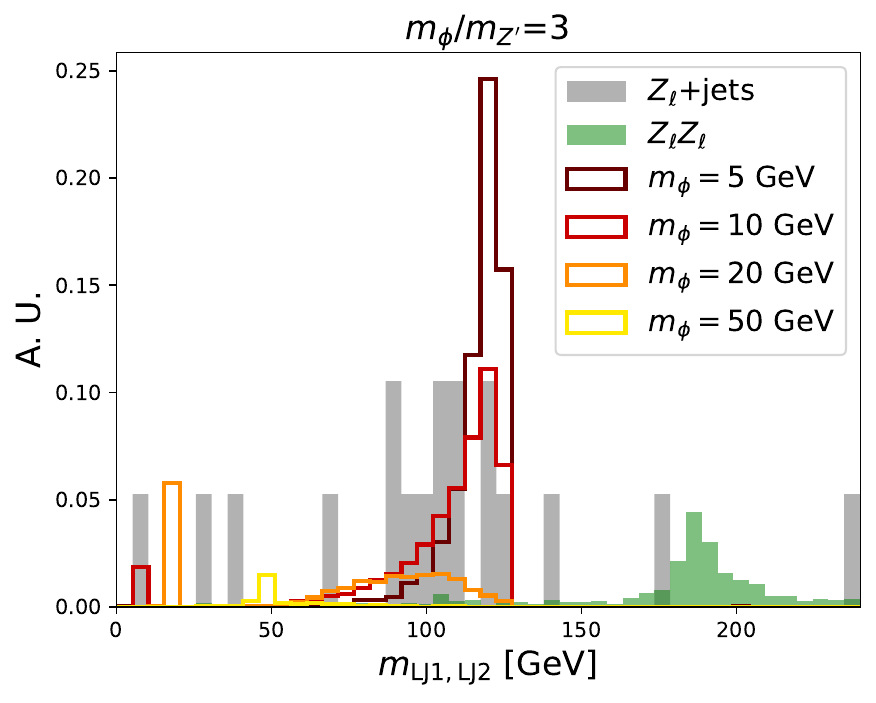}
		\includegraphics[width=0.32\textwidth]{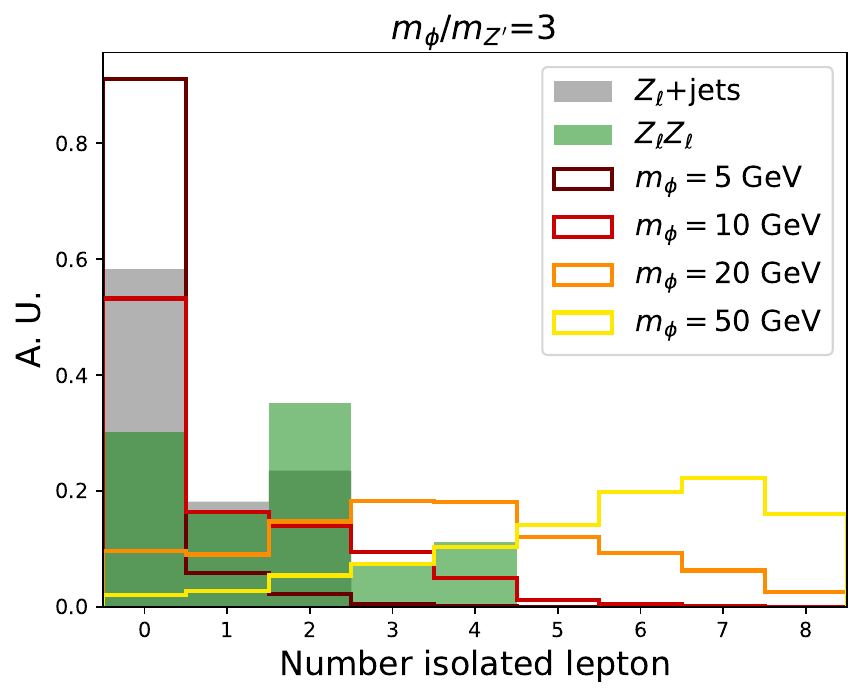}\\
		\includegraphics[width=0.32\textwidth]{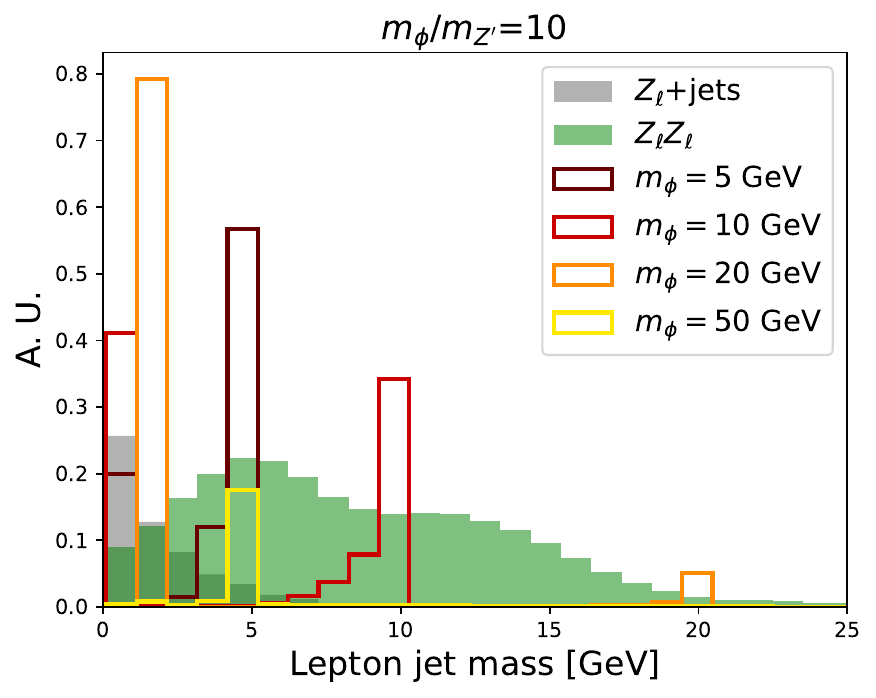}
		\includegraphics[width=0.32\textwidth]{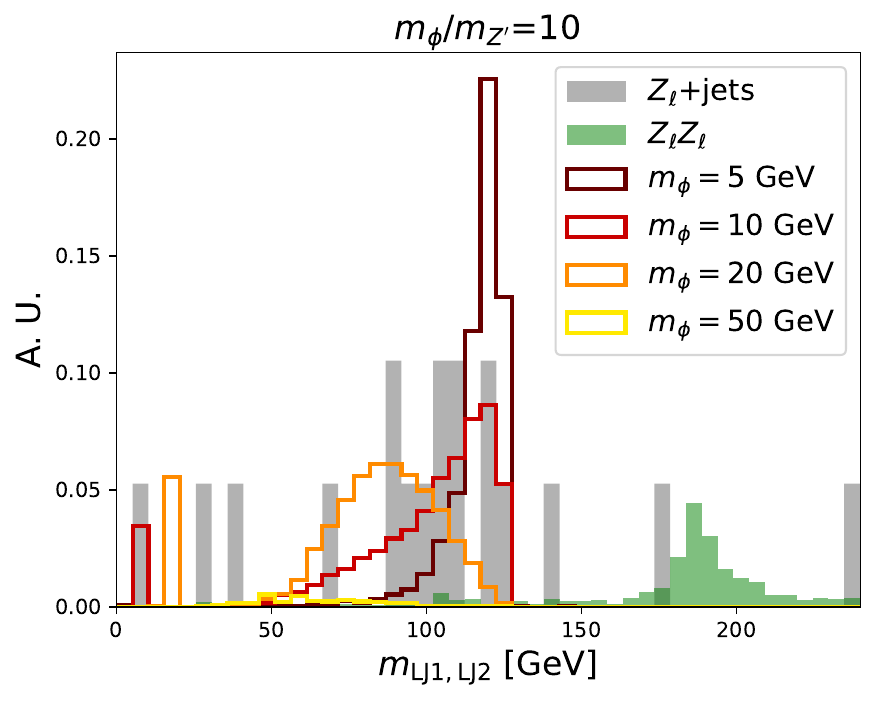}
		\includegraphics[width=0.32\textwidth]{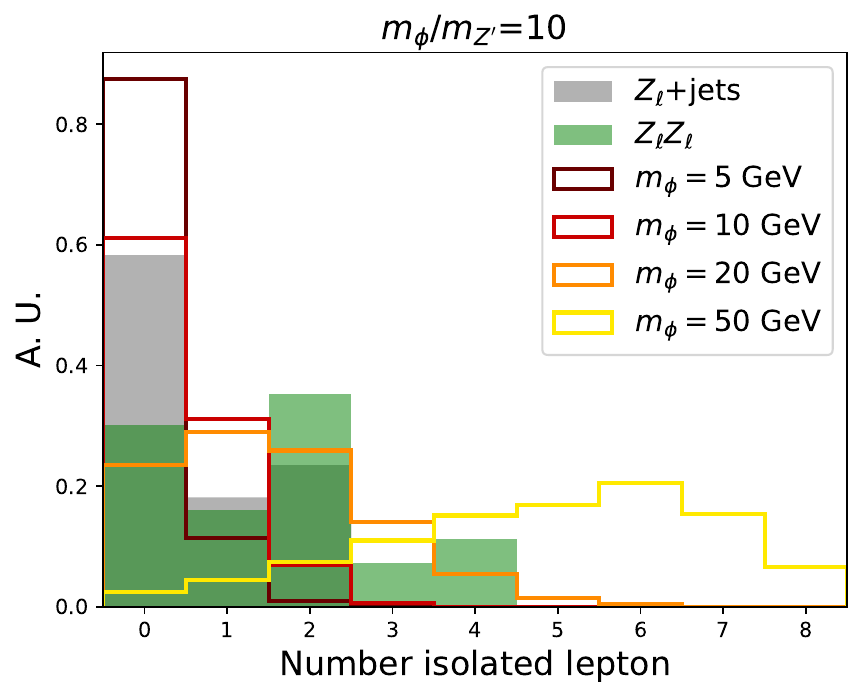}
	\end{center}
\caption{Distributions of the invariant mass of the leading LJ (left panel), the invariant mass of the LJ pair (middle panel) and the number of isolated leptons (right panel) for signals and backgrounds. No selection cut has been applied to the events used in these plots, except for the requirement that each LJ must contain at least two constituent leptons.
\label{fig:vars}}
\end{figure}

To illustrate the sensitivities of the LJ search, two mass ratios $m_{\phi}/m_{Z^\prime}=3$ and 10 are considered. In Fig.~\ref{fig:vars}, we plot the distribution of 
the invariant mass of the leading LJ (left panel), that of two LJs (center panel) and the number of isolated leptons (right panel) for the signal and the background processes. 
Note that for the distribution of the invariant mass of the LJ pair, even though we have simulated $10^7$ leptonic decaying $Z$+jets events, the selected number of events in SRs is still too small to give a smooth distribution. 
Although our LJs contain hadronic constituents, only the momenta of leptonic (including ID track if it is close to an electron) constituents are summed to calculate the LJs invariant mass. This method can also help to alleviate the pileup contamination. We can find that the invariant mass of each LJ peaks at around $m_\phi$ for $m_{\phi} \lesssim 10$ GeV. A larger mass ratio $m_{\phi}/m_{Z^\prime}$ leads to a more remarkable peak. 
Since the LJ pair tends to be produced from the decay of the SM-like Higgs boson, the invariant mass distribution of the LJ pair has a cutoff at around $m_{h} (=125$ GeV), as shown in the middle panel of Fig.~\ref{fig:vars}. On the other hand, the $Z_\ell Z_\ell$ background tends to have a peak at around twice of the $Z$ boson mass. Thus, it can be efficiently suppressed by requring $m_{\text{LJ1}, \text{LJ2}} \lesssim 150$ GeV. 
In the right panel of Fig.~\ref{fig:vars}, we plot the number of isolated leptons for our signal and background processes. 
We see that the number of isolated leptons exceeds four when $m_{\phi} \gtrsim 20$ GeV. A smaller mass hierarchy between $\phi$ and $Z^\prime$ leads to even higher number of isolated leptons.

\begin{figure}[htbp]
	\begin{center}
		\includegraphics[width=0.49\textwidth]{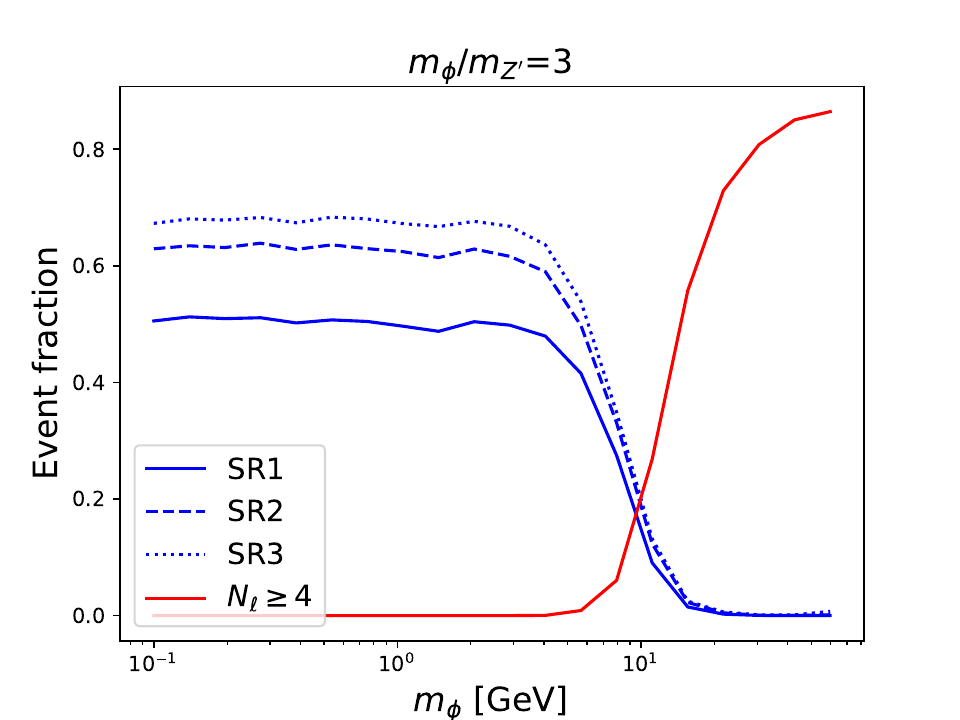}
		\includegraphics[width=0.49\textwidth]{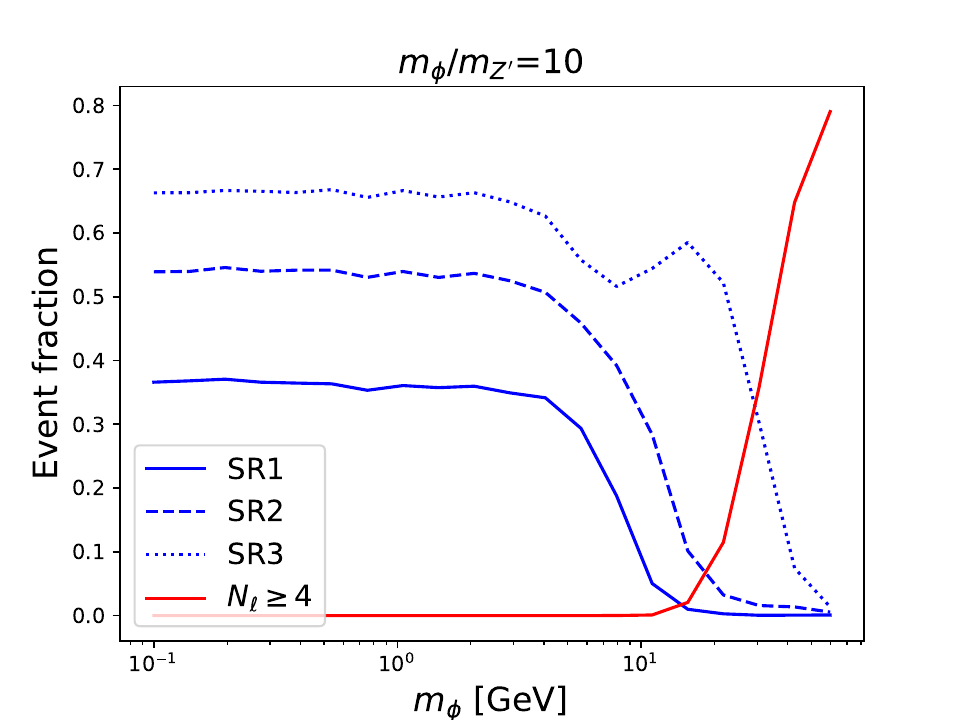}
	\end{center}
\caption{Efficiencies of three SRs (blue curves), as well as at least four isolated leptons (red solid curve) for $m_\phi/m_{Z'}=3$ (left) and 10 (right).
In this plot, we consider the $Z'$ decay into an electron pair. }
\label{fig:frac}
\end{figure}

As we have seen in Fig.~\ref{fig:vars}, the searches for LJs and multi-isolated leptons
play a complementary role to explore both the small and large mass regions.  
In Fig.~\ref{fig:frac}, we show the efficiencies of our SRs and the selection criterion requiring at least four isolated leptons.
The complementary nature of these two searches can clearly be seen.
For the moderate mass ratio $m_\phi /m_{Z^\prime} =3$, the efficiencies of LJs decline significantly when $\phi$ mass exceeds approximately $5$ GeV. On the other hand, the multi-lepton selection exhibits high efficiency for $m_{\phi} \gtrsim 20$ GeV. 
As for high mass ratio $m_\phi /m_{Z^\prime} =10$, the overall efficiencies of our SRs decrease, while the decline in efficiency occurs at larger $m_{\phi}$, especially for SR3. 
Meanwhile, the multi-lepton search requires $m_{\phi} \gtrsim 30$ GeV to achieve an efficiency exceeding approximately 0.5. 

\begin{figure}[htbp]
	\begin{center}
		\includegraphics[width=0.49\textwidth]{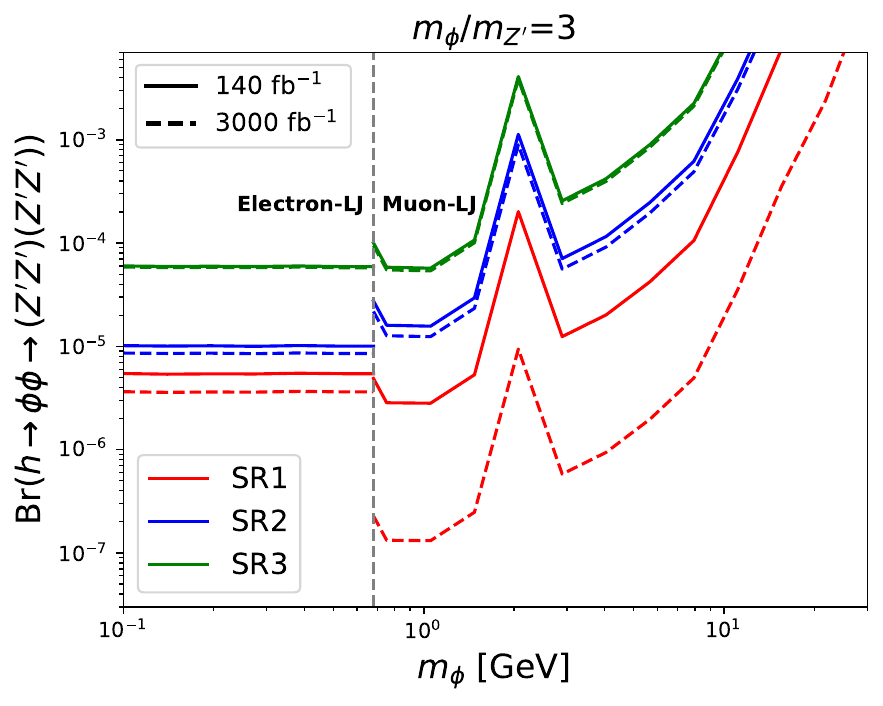}
  \includegraphics[width=0.49\textwidth]{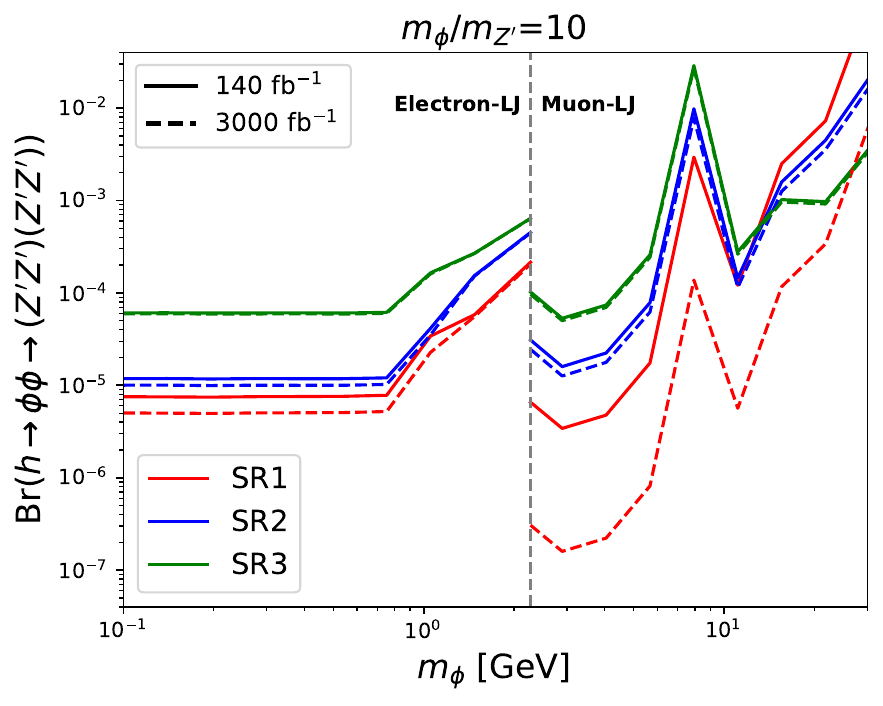}
	\end{center}
\caption{Exclusion limits on the branching ratio of $h \to \phi \phi \to (Z^\prime Z^\prime) (Z^\prime Z^\prime)$ obtained from different SRs in two benchmark scenarios. In the mass region $m_{Z^\prime} > 2 m_\mu$, only the LJs which contain at least two muonic constituents are countted. } 
\label{fig:Brlimit}
\end{figure}

We can calculate the exclusion limits on the Higgs decay branching ratio $h\to \phi \phi \to (Z^\prime Z^\prime) (Z^\prime Z^\prime)$ 
by using the cross sections of the Higgs production~\cite{Anastasiou:2016cez}, the $ZZ$ production at the next-to-leading-order in QCD~\cite{Campbell:2011bn} and the measured $Z_\ell$+jets production~\cite{ATLAS:2024irg}, together with the signal and background selection efficiencies as have been discussed above. 
The limit is calculated at a given integrated luminosity ${\cal L}$ by 
\begin{align}
\frac{\sigma_S \epsilon_S \times \sqrt{\mathcal{L}}}{\sqrt{ \sigma_{Z_\ell Z_\ell} \epsilon_{Z_\ell Z_\ell}  +  \sigma_{Z_{\ell}\text{+jets}} \epsilon_{Z_{\ell}\text{+jets}} + [0.05\times (\sigma_{Z_\ell Z_\ell} \epsilon_{Z_\ell Z_\ell}  +  \sigma_{Z_{\ell}\text{+jets}} \epsilon_{Z_{\ell}\text{+jets}})]^2 \mathcal{L}  }}=2~,
\end{align}
where $\sigma_i$ and $\epsilon_i$ respectively represent the production cross section and the efficiency for each event $i$ $(i=S,~Z_{\ell}\text{+jets},~Z_\ell Z_\ell$) with $``S"$ denoting the signal.
The signal cross section is estimated by 
\begin{align}
\sigma_S^{} = \sigma_h \times \text{BR}(h \to 4Z') \times [\text{BR}(Z' \to \ell^+\ell^-)]^4.  
\end{align}
We assume the systematic uncertainty to be 5\%~\cite{ATLAS:2024zxk}. 
The definition of the factor of $[\text{BR}(Z'\to \ell^+\ell^-)]^4$ is changed depending on the mass of $Z'$ as follows:
\begin{align}
&[\text{BR}(Z'\to \ell^+\ell^-)]^4\notag\\
&=\begin{cases}
    \Big[[\text{BR}(Z^\prime \to \mu \mu)]^2+2 
    \text{BR}(Z^\prime \to \mu \mu) \text{BR}(Z^\prime \to ee)\Big]^2~~\text{for}~~m_{Z'} \geq 2m_\mu \\
    [\text{BR}(Z^\prime \to ee)]^4~~\text{for}~~m_{Z'} < 2m_\mu  .  
\end{cases}
\label{eq:br}
\end{align}
The branching ratio of $Z'$ depends on the model. We here consider the dark photon case as an example, which can be realized by taking $\tilde{X}_\Psi = 0$ appearing in Eq.~(\ref{eq:x-charge}), except for $\Psi =\Phi$. 
We then estimate the branching ratios by applying the method in Ref.~\cite{Ilten:2018crw} with data of $\sigma(e^+e^- \to {\rm hadron})/\sigma(e^+e^- \to \mu^+\mu^-)$ from the {\tt DARKCAST} code. 
For instance, the factor of $[\text{BR}(Z'\to \ell^+\ell^-)]^4$ is given to be around 0.23 and 0.0046 for $m_{Z'}=0.5$ GeV and 10 GeV, respectively. 
For $m_{Z'} <  2m_\mu$, $[\text{BR}(Z'\to \ell^+\ell^-)]^4$ is almost unity, because the dark photon cannot decay into a neutrino pair.

The results are presented in Fig.~\ref{fig:Brlimit}. 
The SR1 provides the best sensitivities for all cases, except for the simultaneous heavy $\phi$ and large mass ratio. 
Despite the suppression due to the branching ratio mentioned above, 
the muonic LJ signature in the mass region $m_{Z^\prime}>2 m_\mu$ can exhibit better sensitivity than the electronic LJ signature in the mass region $m_{Z^\prime}<2 m_\mu$ due to lower background levels.  
Notice that the bumps at $m_{Z'} \sim 1$ GeV for the sensitivity curves are due to the resonant enhancement of the hadronic branching ratios that suppresses leptonic ones.
We also find that the improvement in sensitivity reaches is mild with increasing integrated luminosity for the SR2 and SR3, where the systematic uncertainty is dominant.

\section{Constraints and sensitivity to model parameters \label{sec:4}}

In this section, we estimate the sensitivity to model parameters such as the new scalar boson mass $m_\phi$ and the scalar mixing angle $\alpha$, combining the constraints on BR$(h \to \phi \phi \to Z'Z'Z'Z')$ derived from the LJ analysis in the previous section and our formulas for the decay rates in Eqs.~\eqref{decay:hphiphi} and \eqref{eq:phi-decay-zpzp}. 
We also compare this new constraint with that given by the existing flavor data and the $h \to Z'Z'$ decay. 
As in the previous section, we consider the dark photon case as an example of the concrete model. 

For the smaller scalar mass region ($m_\phi \lesssim 5$ GeV), severe constraints have been taken on the scalar mixing $\alpha$ from searches for rare meson decays into final states with $\phi$~\cite{BNL-E949:2009dza,MicroBooNE:2021sov,MicroBooNE:2022ctm,KOTO:2020prk,NA62:2020pwi,NA62:2021zjw,Gorbunov:2021ccu,CHARM:1985anb,Winkler:2018qyg,Belle-II:2023ueh,BaBar:2015jvu,LHCb:2015nkv,LHCb:2016awg}. 
See Ref.~\cite{Ferber:2023iso} for the review of such constraints.
For the larger scalar mass region ($m_\phi > 10$ GeV),  there are several constraints from 
direct searches for a neutral scalar boson at LHC~\cite{CMS:2018zvv,ATLAS:2021hbr}. 
In addition, we consider the constraint from the search for the Higgs boson decay $h \to Z'Z'$ for $m_{Z'} < m_h/2$, where the upper bound on BR$(h \to Z'Z')$ has been given in Ref.~\cite{ATLAS:2024zxk} for the dark photon case.
The upper bound on $|\sin\alpha|$ can then be estimated from that of BR$(h \to Z'Z')$ applying our formula of the decay rate in Eq.~\eqref{eq:hzpzp}.

\begin{figure}[t!]
\begin{center}
\includegraphics[width=0.49\textwidth]{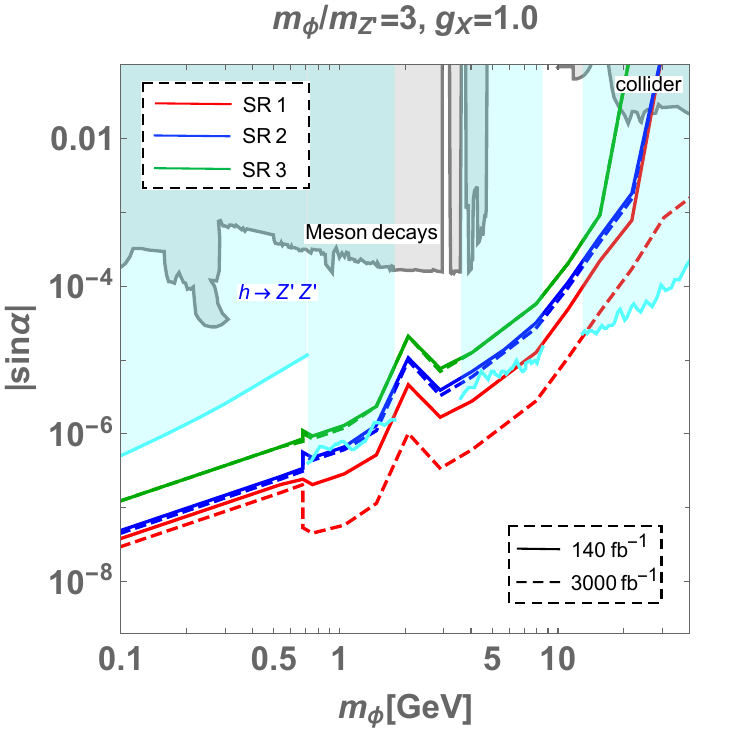}
\includegraphics[width=0.49\textwidth]{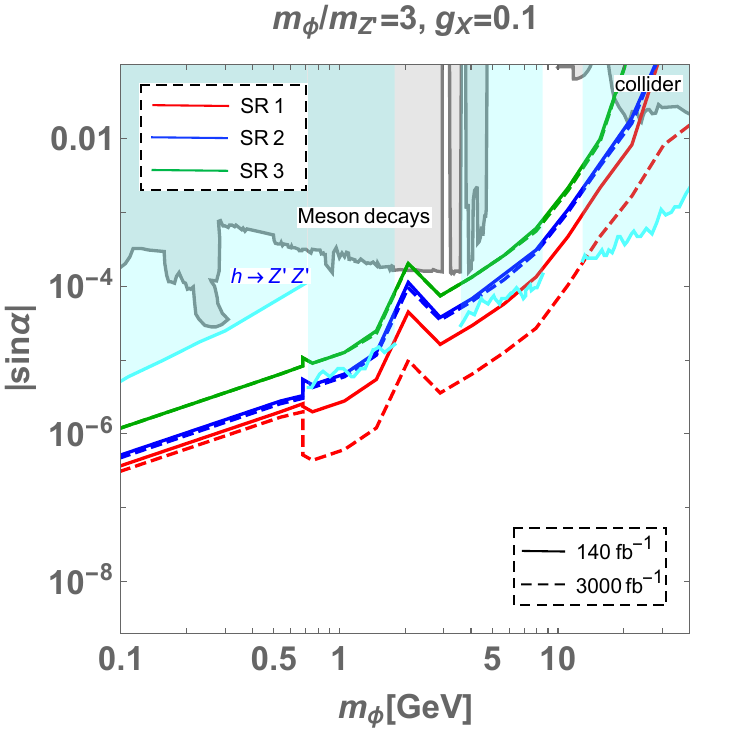} \\
\includegraphics[width=0.49\textwidth]{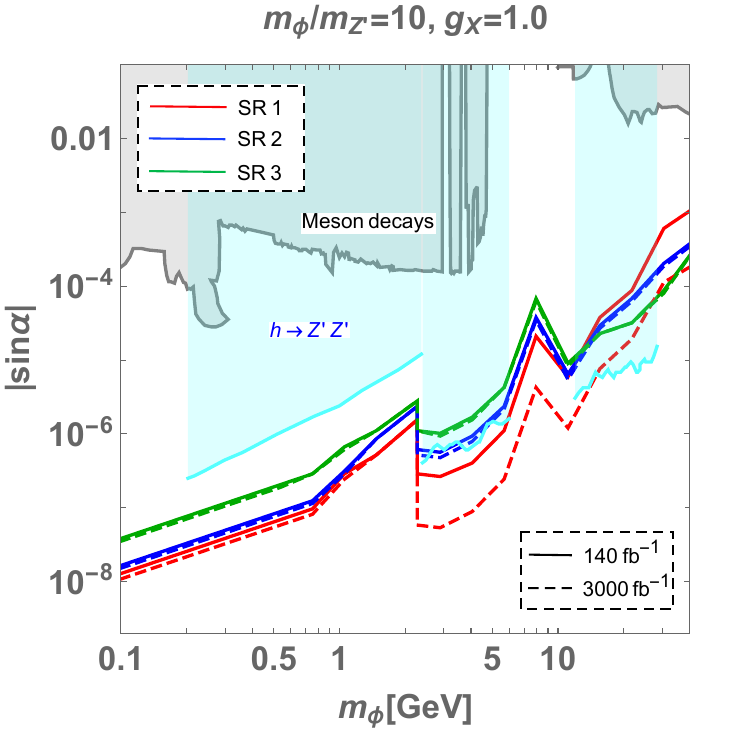}
\includegraphics[width=0.49\textwidth]{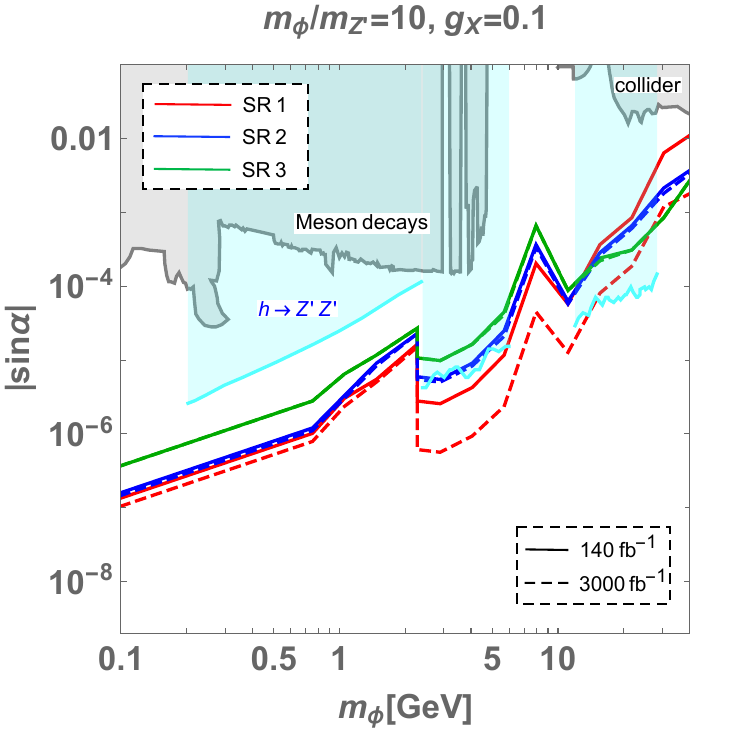}
\end{center}
\caption{Solid (Dashed) curves indicate 95\% C.L. exclusion limits on the scalar mixing $|\sin\alpha|$  in the dark photon case 
for integrated luminosity 140 (3000) fb$^{-1}$. 
The red, blue and green curves respectively represent the results using SR1, SR2 and SR3 which are defined in Sec.~\ref{sec:3}. 
} 
\label{fig:limit-parameters}
\end{figure}

In Fig.~\ref{fig:limit-parameters}, we show the upper limit on $|\sin\alpha|$ at 95\% C.L. as a function of $m_\phi$, where the solid (dashed) curves correspond to the limits with the integrated luminosity 140 fb$^{-1}$ (3000 fb$^{-1}$), while the red, blue and green curves denote the results using the SR1, SR2 and SR3, respectively. 
We indicate the values of the ratio $m_\phi/m_{Z'}$ and the new gauge coupling $g_X^{}$ on the top of each panel. 
The region excluded by current experimental constraints are also shown as the gray-shaded area (meson decays and collider experiments) and the cyan-shaded area (search for the $h \to Z'Z'$ decay).
Note that both the limits by $h \to Z'Z'Z'Z'$ and $h \to Z'Z'$ become weaker when the gauge coupling $g_X^{}$ decreases since the corresponding decay widths are scaled as $\Gamma(h \to Z'Z'/\phi \phi) \propto v_\Phi^{-2} \simeq g_X^2 X_\Phi^2/m_{Z'}^2$ as shown in Eqs.~\eqref{eq:hzpzp} and \eqref{decay:hphiphi}. 
We find that the strongest upper bound on $|\sin \alpha|$ is obtained by our LJ analysis when the scalar mass is $m_\phi \lesssim 5$ GeV. In particular, we can realize the high sensitivity by applying our SR1 scheme of selecting events.
For the larger scalar mass region, the limit from $h \to Z'Z'$ process tends to be stronger since leptons are less collimated for heavy $\phi$ and $Z'$ masses reducing the effectiveness of our selection. Therefore, we expect higher sensitivity by combining the searches for $Z'Z'$ and $Z'Z'Z'Z'$ modes from the rare decay of the SM-like Higgs boson.  

\section{Conclusions \label{sec:5}}

We have studied an exotic decay of the 125 GeV Higgs boson ($h$) into four $Z'$ via the decay chain of $h \to \phi\phi \to Z'Z'Z'Z'$ with $\phi$ being a new scalar boson whose vacuum expectation value spontaneously breaks a new $U(1)$ gauge symmetry. 
The charged leptons generated from the $Z'$ decay tend to form lepton-jets rather than isolated leptons when the $\phi$ and $Z'$ masses are typically smaller than about 10 GeV, because of their boosted effects.   
Therefore, the Higgs boson can decay into final states with multi-lepton jets in our scenario. 

We have performed the simulation studies for signal and background events including detector level effects, in which we have introduced three signal regions (SRs) requiring different number of constituent charged leptons in the lepton-jets. 
We have found that the SR1 which requires two lepton-jets including at least three constituent leptons gives the highest signal to background ratio.  
Applying the SR1, we have obtained the constraint BR$(h \to Z'Z'Z'Z') < 10^{-6}~(10^{-7})$ for $m_{Z'}\simeq 1$ GeV by using the muonic-lepton jets assuming the integrated luminosity of 140 fb$^{-1}$ (3000 fb$^{-1}$) at LHC. 
This bound was obtained by assuming the dark photon case, but similar bounds can also be obtained in other models with an additional $U(1)$ gauge symmetry, because the decay branching ratio of $Z'$ does not change so much from the dark photon case.   
For lighter $Z'$ ($< 2m_\mu$), we have used the electronic lepton-jets instead of the muonic ones, by which the upper limit on the branching ratio is obtained to be of order $10^{-6}$-$10^{-5}$.

Finally, we have applied the bound on the branching ratio of $h$ to the constraint on model parameters such as a mixing angle between $h$ and $\phi$.  It has been shown that stronger bounds on the mixing angle are obtained in the dark photon case as compared with those given by flavor constraints and the Higgs decay $h \to Z'Z'$ in the mass range of $m_{Z'}\lesssim 10$ GeV. 

Let us give a brief comment on the other models with a $U(1)$ gauge symmetry such as $U(1)_{B-L}$~\cite{Pati:1973uk,Davidson:1978pm,Marshak:1979fm} and $U(1)_{L_i - L_j}$~\cite{Foot:1990mn,He:1990pn}. 
In these models, a new gauge coupling constant $g_X^{}$ has been highly constrained to be typically smaller than ${\cal O}(10^{-4})$ by the experimental data, e.g., flavor constraints~\cite{Bauer:2018onh}. In these cases, the upper limit on $|\sin\alpha|$ becomes 4 orders of magnitude weaker than that shown in Fig.~\ref{fig:limit-parameters}, and the constraint becomes comparable with that given by the meson decays.

To conclude, we have shown that the search for lepton-jets is quite useful to explore the $Z'$ boson, especially the dark photon, via the decay of the Higgs boson.

\section*{Acknowledgments}
The work was supported by the Fundamental Research Funds for the Central Universities (T.~N.,J.~L.), and by the Natural Science Foundation of Sichuan Province under grant No.~2023NSFSC1329 and the National Natural Science Foundation of China under grant Nos.~11905149 (J.~L.). 
\\\\
{\it Note added:} \\
After this work was almost completed, Ref.~\cite{Cheng:2024gfs} appeared, where the bound on the cross-section of the $h \to \phi\phi \to (Z^\prime Z^\prime) (Z^\prime Z^\prime) \to (4\ell) (4\ell)$ process has been studied by using the existing multi-lepton searches at the ATLAS experiment~\cite{ATLAS:2023jyp} with $m_\phi \gtrsim 5$ GeV. 
It has been found that the bound on the production cross section is around a few femto-barn when $m_{Z'} \sim \mathcal{O}(10)$ GeV. 
Our analyses using LJs mainly focused on the case with $m_{Z'} < 5$ GeV.


\bibliography{references}
\end{document}